\pdfoutput=1
\documentclass[a4paper, 11pt]{article}

\usepackage[utf8]{inputenc}
\usepackage[T1]{fontenc}
\usepackage[american]{babel}

\usepackage[lining, scale=0.88]{FiraMono}
\usepackage[book, semibold, lining, tabular, scale=0.88]{FiraSans}[2018/05/23]
\usepackage[charter, sfscaled=false, ttscaled=false, euro=false]{mathdesign}
\usepackage[scaled=0.96, lining]{XCharter}
\usepackage{mweights}

\usepackage[sf, bf, raggedright, explicit]{titlesec}
\usepackage[style]{abstract}

\usepackage[max6,affil-it]{authblk}

\usepackage{amsmath}
\usepackage{url,nicefrac,siunitx,xspace,textcomp,color,xcolor,setspace,comment,cite}
\definecolor{darkgreen}{cmyk}{0.9, 0, 0.9, 0}
\definecolor{darkred}{cmyk}{0, 0.6, 0.6, 0}
\usepackage[labelformat=simple,list=true]{subcaption}

\usepackage[
  a4paper,
  width=0.8\paperwidth,
  height=0.76\paperheight,
]{geometry}

\usepackage{etoolbox}
\apptocmd{\thebibliography}{\setlength{\itemsep}{0pt}}{}{}

\usepackage{hyperref}
\hypersetup{
  colorlinks=true,
  linkcolor=blue,   % internal: \ref, \autoref, table of contents
  citecolor=darkgreen,   % \cite
  urlcolor=blue,     % \href and \url
  citebordercolor=0 1 0,
  filebordercolor=0 1 1,
  linkbordercolor=1 0 0,
  menubordercolor=1 0 0,
  urlbordercolor=1 0 0,
  runbordercolor=0 .7 .7,
}
\usepackage{orcidlink}
\usepackage{graphicx}%
\usepackage{multirow}%
\usepackage[title]{appendix}%
\usepackage{textcomp}%
\usepackage{manyfoot}%
\usepackage{booktabs}%
\usepackage{algorithm}%
\usepackage{algorithmicx}%
\usepackage{algpseudocode}%
\usepackage{listings}%

\DeclareGraphicsExtensions{.pdf,.png,.jpg,.eps}
\graphicspath{{./}{./figs/}}

\newcommand{\be}{\begin{equation}}
\newcommand{\ee}{\end{equation}}
\newcommand{\tev}{\mbox{TeV}\xspace}
\newcommand{\gev}{\mbox{GeV}\xspace}
\newcommand{\ifb}{\mbox{fb$^{-1}$}\xspace}
\newcommand{\gd}{\ensuremath{g_{\mathrm D}}\xspace}
\newcommand{\hf}{\nicefrac{1}{2}\xspace}
\newcommand{\bbi}{\ensuremath{\beta_\text{BI}}\xspace}
\newcommand{\mgg}{\ensuremath{m_{\gamma\gamma}}\xspace}
\newcommand{\qeff}{\ensuremath{Q_\text{eff}}\xspace}

\newcommand{\papertitle}{Constraining magnetic monopoles and multiply charged particles with diphoton events at the LHC}

\title{\papertitle}
\author[1]{Vasiliki A.\ Mitsou\,\orcidlink{0000-0002-1533-8886}\thanks{vasiliki.mitsou@ific.uv.es} }
\author[1,2]{Emanuela Musumeci\,\orcidlink{0000-0002-2077-4949}\thanks{Corresponding author. emanuela.musumeci@cern.ch} }
\affil[1]{Instituto de F\'isica Corpuscular (IFIC), CSIC -- Universitat de Val\`encia, C/ Catedr\'atico Jos\'e Beltr\'an 2, 46980 Paterna (Valencia), Spain}
\affil[2]{Department of Physics and Astronomy, University of Alabama, Tuscaloosa, 35487, Alabama, USA}

\date{}  %\date{\today}

\begin{document}

\maketitle

\begin{abstract}
\noindent The LHC is achieving energies never reached before directly at colliders, opening up possibilities for the discovery of exotic particles in the TeV mass range. Such states include magnetic monopoles,  which can explain the electric charge quantisation and restore the symmetry in Maxwell's equations with respect to the magnetic and electric fields. Scenarios proposed to shed light to dark matter and neutrino masses introduce high-electric-charge objects (HECOs). The existence of both classes of particles can be probed in precision measurements in a manner complementary to direct searches. We focus on the contributions of such virtual  particles to light-by-light scattering in the context of effective field theories and a Born--Infeld scenario. Specifically, measurements of central exclusive production of photon pairs with proton tagging carried out by the CMS-TOTEM Precision Proton Spectrometer with LHC Run~2 proton--proton collision data are used to constrain magnetic monopole and HECOs. Resummation techniques have been employed to deal with the large HECO-photon coupling. Masses of up to a few tens of TeV have been excluded for monopoles and HECOs of various spins and magnetic and electric charges, respectively.
\\\\
\textbf{Keywords:} magnetic monopole, high-electric-charge object, light-by-light scattering, effective field theory, Born Infeld, diphoton, proton tagging, resummation
\end{abstract}

%%%%%%%%%%%%%%%%%%%%%%%%%%%%%%%%%%%%%%%%%%%%%%%%%%%%%%%%%%%%
\section{Introduction}\label{sc:intro}

Numerous searches for magnetic monopoles (MMs) have been carried out~\cite{Mavromatos:2020gwk,Mitsou:2025ufr,Mitsou:2026zcf} since Paul Dirac predicted their existence in 1931~\cite{Dirac:1931kp}. He proved that MMs could explain the discrete nature of the electric charge, providing the so-called Dirac Quantisation Condition (DQC) 
\begin{equation}
   \alpha g= \frac{n}{2}  e, \quad n \in \mathbb{Z}, \label{dqc}
\end{equation}
where $\alpha=e^2/(4 \pi)$ is the fine structure constant, \textit{e} is the elementary electric charge and \textit{g} is the MM charge in natural units, $\hbar=c=1$, in which we work throughout this study. In Dirac’s formulation, monopoles are assumed to be point-like particles and quantum mechanical consistency conditions lead to \eqref{dqc}, establishing the value of their magnetic charge as multiples of the Dirac charge $\gd = e/(2\alpha)$. Mass $M$ and spin $s$ are free parameters of the theory. In 1974 't Hooft and Polyakov showed that grand unified theories of strong and electroweak interactions predicted the existence of MM as a soliton with spontaneous symmetry breaking~\cite{tHooft:1974kcl,Polyakov:1974ek}. A comprehensive review of theories predicting MMs is given in Ref.~\cite{Mavromatos:2020gwk}.
An interesting feature of MMs is the symmetrisation of Maxwell's equations with respect to magnetic and electric fields, leading to the \textit{electric--magnetic duality}~\cite{Dirac:1931kp,Dirac:1948um,Milton:2006cp}. This way, the monopole--photon  coupling may be $\beta$-dependent, with $\beta= \sqrt{1-\frac{4M^2}{\hat{s}}}$, with $\hat{s}$ the Mandelstam variable~\cite{Schwinger:1976fr,Milton:2006cp,Epele:2012jn}. In this work, we consider point-like, i.e.\ structureless, MMs with $\beta$-independent MM--$\gamma$ couplings. A justification for $\beta$-independence within a framework of effective gauge field theory approach to MM--matter scattering is given in Ref.~\cite{Alexandre:2026auj}.

Due to the electric--magnetic duality, monopoles are expected to be produced in electromagnetic processes analogous to those producing pairs of electrically charged particles. Hypothetical high-electric-charge objects (HECOs) have been proposed, such as Q-balls~\cite{Coleman:1985ki,Kusenko:1997si}, micro-black-hole remnants~\cite{Koch:2005ks,Hossenfelder:2005ku}, quirks~\cite{Kang:2008ea}, scalars in radiative-neutrino-mass models~\cite{R:2020odv,Hirsch:2021wge}, and aggregates of $ud$-~\cite{Holdom:2017gdc} or $s$-quark (strangelets) matter~\cite{Farhi:1984qu}, among others. Depending on the specific scenario, their existence can provide an explanation to the nature of dark matter, the hierarchy problem and the non-vanishing neutrino masses.

MM and HECO pairs can be produced directly at colliders via a Drell--Yan-like photon exchange ---and $Z$ exchange in the case of HECOs--- and through photon-fusion~\cite{Baines:2018ltl,Song:2021vpo}, whilst monopole-antimonopole pairs may be also created in strong magnetic fields via the Schwinger mechanism~\cite{Affleck:1981ag,Gould:2017fve,Gould:2019myj, Rajantie:2024wuw}. Searches for both classes of (meta)stable objects are predominantly based on high ionisation energy loss and on large time-of-flight measurements characterising heavy slow-moving particles. Such searches are being conducted exploiting all three aforementioned production processes at the Large Hadron Collider (LHC)~\cite{Mavromatos:2020gwk,Mitsou:2024tyq,Mitsou:2025ufr, Mitsou:2026zcf} by the general-purpose ATLAS experiment~\cite{ATLAS:2012bda,ATLAS:2015tyu,ATLAS:2019wkg,ATLAS:2023esy,ATLAS:2024nzp} and the dedicated MoEDAL detectors~\cite{MoEDAL:2016jlb,MoEDAL:2016lxh,MoEDAL:2017vhz,MoEDAL:2019ort,MoEDAL:2020pyb,MoEDAL:2021vix,MoEDAL:2021mpi,MoEDAL:2023ost,MoEDAL:2024wbc} with good prospects for the future~\cite{Hirsch:2021wge,Altakach:2022hgn,Maselek:2025cfm}.

\begin{figure}[ht] 
\centering
\includegraphics[width=0.4\textwidth]{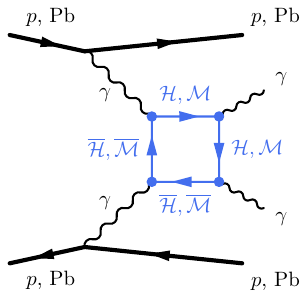}
   \caption{HECO, $\mathcal{H}$, and magnetic monopole, $\mathcal{M}$, contributions to light-by-light scattering.}
    \label{fig:box}
\end{figure}

Monopoles may be probed \emph{indirectly} via the formation of a monopole--antimonopole bound state, the \emph{monopolium}~\cite{Dirac:1931kp,Hill:1982iq,Vento:2007vy} and its subsequent decay to two or more photons~\cite{Epele:2012jn,Barrie:2021rqa,daSilva:2023jxd} or their excited states~\cite{Fanchiotti:2022xvx,Fanchiotti:2023jmx}. Likewise, HECOs may be constrained through diphoton-event measurements as they can also form a bound state~\cite{Altakach:2022hgn}. Moreover, MMs and HECOs can contribute to ``box diagrams'' of light-by-light (LbL) scattering such as the one in Fig.~\ref{fig:box}, which may lead to an enhancement of multiphoton production in colliders. This work focuses precisely on this process, assuming a $\beta$-independent $\gamma$--MM coupling~\cite{Musumeci:2025tuw}. 

This article is structured as follows. Section~\ref{sc:cep} introduces the reader to the central exclusive $\gamma\gamma$ production and the experimental results used subsequently to constrain physics beyond the Standard Model (BSM). The framework of effective field theories (EFTs) is discussed in Section~\ref{sc:eft}, whereby leads to indirect mass limits for HECOs and MMs in  Sections~\ref{sc:heco} and~\ref{sc:mm-eft}, respectively. For a specific MM scenario, additional bounds are estimated in the Born--Infeld theory in Section~\ref{sc:born}. The conclusions are summarised in Section~\ref{sc:conclu}, where future prospects are also provided. 

%%%%%%%%%%%%%%%%%%%%%%%%%%%%%%%%%%%%%%%%%%%%%%%%%%%%%%%%%
\section{Central exclusive diphoton production }\label{sc:cep}

The final-state signature of interest involves diphoton production through central exclusive processes (CEP):
\begin{center}
    $A + B \xrightarrow{\gamma  \gamma} A   \gamma  \gamma  B$.
\end{center}
Here $A, B = p, \text{Pb}$ are the incoming hadrons or nuclei which undergo scattering at an extremely small angle relative to the beam. The resulting final state consists of two photons, that can be detected by the central detector, and two hadrons scattered are small angles. LbL scattering in CEP is a well-suited channel to study possible contributions from new particles~\cite{dEnterria:2013zqi}. 
%The large photon-photon luminosity and the high invariant mass range make two-photon collisions at colliders also of interest for the search for new particles and new physics. This includes the possible production of the
%Higgs-boson in the $\gamma \gamma$-production channel or new physics beyond the standard model.\cite{Baur_2002}

Due to the low virtualities $Q^2 < 1/R^2$, where $R$ is the charge radius, the photons exchanged during collisions between hadrons are almost on-shell. Using the equivalent photon approximation (EPA) framework~\cite{Budnev:1975poe}, the elastic diphoton production cross section in the collision of hadrons $A$ and $B$ can be factorised. This factorisation involves the elementary cross section for $\gamma \gamma \to \gamma \gamma$ at $\sqrt{s_{\gamma \gamma }}$, convoluted with the EPA spectra from the two colliding beams: 
\begin{equation}
 \sigma^\text{excl}_{\gamma \gamma \to \gamma \gamma}= \int d\omega_1 d \omega_2 \frac{f_{\gamma/A}(\omega_1)}{\omega_1} \frac{f_{\gamma/B}(\omega_2)}{\omega_2} \sigma_{\gamma\gamma \to \gamma \gamma} (\sqrt{s_{\gamma \gamma }}), \label{xs_excl}
\end{equation}
where $\omega_1$ and $\omega_2$ represent the photon energies and ${f_{\gamma/A,B}(\omega)}$ are the photon fluxes at energy $\omega$  emitted by the hadrons $A$ and $B$, respectively. This approach treats all charged particles producing electromagnetic fields at high energies as photon beams.

The centre-of-mass energy of the diphoton system is given by $\sqrt{s_{\gamma \gamma }} = \mgg = \sqrt{ 4 \omega_1 \omega_2}$. The maximum value for $\omega$ occurs when $\omega^\text{max} \simeq  \gamma/b_\text{min}$,  
where $b_\text{min}$ is the minimum separation between the two charges of radius $R_{A,B}$ and $\gamma=\frac{\sqrt{s_\text{NN}}}{2m_\text{N}}$ is the beam Lorentz factor. Here, $\sqrt{s_\text{NN}}$ is the nucleon-nucleon centre-of-mass energy and $m_\text{N}$ is the nucleon mass~\cite{dEnterria:2013zqi}.

LbL scattering in either heavy-ion collisions~\cite{ATLAS:2020hii,CMS:2018erd,CMS:2024bnt} or in proton collisions~\cite{Baldenegro:2018hng,ATLAS:2023zfc} have already been used to search for BSM states, such as axion-like particles. Such searches for resonant diphoton production can also constrain other exotic states such as monopolia~\cite{Vento:2007vy, Epele:2012jn,Barrie:2021rqa}.

In the case of proton--proton ($pp$) collisions, a more energetic photon spectrum can be generated than heavy-ion collisions, enabling to probe much higher diphoton invariant masses. Larger datasets of $pp$ collisions are available in the LHC experiments compared to heavy-ion runs. On top of that, the \textit{proton tagging method} allows the measurement of the photon pair in the central detector while tagging the scattered intact protons using specialised forward proton detectors~\cite{Fichet:2013gsa,Baldenegro:2018hng}.

In the LHC Run~2, two dedicated detectors, the ATLAS Forward Physics (AFP)~\cite{Adamczyk:2015cjy} and the CMS-TOTEM Precision Proton Spectrometer (CT-PPS)~\cite{CMS:2014sdw}, were installed symmetrically with respect to the respective interaction points to explore CEP. Searches for exclusive diphoton production using forward proton tagging have been performed by CT-PPS~\cite{TOTEM:2021zxa,TOTEM:2023ewz} and by AFP-ATLAS~\cite{ATLAS:2023zfc}. In this work we use the latest CMS-TOTEM results to constrain MMs and HECOs. That analysis required the detection in the same event of two back-to-back photons with large transverse momenta by the CMS lead-tungstate-crystal electromagnetic calorimeter and of two opposite-side forward protons in PPS~\cite{TOTEM:2023ewz}. Among other selection criteria imposed to suppress the dominant inclusive photon background, large diphoton invariant mass is required, resulting in sample of events with $350~\gev < \mgg \lesssim 2~\tev$. No exclusive $\gamma \gamma$ event was found above expected backgrounds and limits were set by CMS-TOTEM on anomalous quartic photon couplings allowing constraining BSM scenarios, as we shall see in Section~\ref{sc:eft}. 

%%%%%%%%%%%%%%%%%%%%%%%%%%%%%%%%%%%%%%%%%%%%%%%%%%%%%x
\section{Effective field theory interpretation}\label{sc:eft}

If the BSM mass scale is heavier than the experimental energy, four-photon interactions can be described by two pure-gauge operators in the following effective Lagrangian~\cite{Gupta:2011be,Fichet:2014uka}
 \begin{equation}
\mathcal{L}_{4\gamma}=\zeta_1 F_{\mu \nu} F^{\mu \nu} F_{\rho \sigma}F^{\rho \sigma}+\zeta_2 F_{\mu \nu} F^{\nu \rho} F_{\rho \lambda}F^{\lambda \mu}, \label{EFT}
\end{equation}
where $F_{\mu\nu}$ is the electromagnetic field strength tensor. These dimension-8 operators are highly suppressed in the Standard Model (SM) \cite{Fichet:2013gsa} and the effect of any object beyond the SM heavier than $M_{\gamma\gamma}$ on the $\gamma \gamma \to \gamma \gamma$ process can be parameterised in terms of the $\zeta_1$, $\zeta_2$ parameters.

As in any other effective theory, this EFT approach is valid in a specific kinematic regime of particle contributions to diphoton production. The particles participating in the loop shown in Fig.~\ref{fig:box} are assumed to be at the decoupling limit, i.e., their masses are much larger than the diphoton invariant mass, $M \gg \mgg$, such that can be treated as point-like~\cite{Ellis:2022uxv}. For lighter particles, the EFT computation needs to be corrected; this can be done by using ad-hoc form factors, as often done in the literature~\cite{Fichet:2014uka}, although the more correct approach is to compute the full one-loop amplitude explicitly.

Moreover, EFTs are non-renormalisable, and unitarity breaks down at high energies. Partial wave analysis imposes conditions between $\zeta_{1,2}$ and \mgg to guarantee that the theory remains unitary~\cite{Ginzburg:1999ej}. Since the bulk of the recorded diphoton events in~\cite{TOTEM:2023ewz} have $\mgg \lesssim 1~\tev$ with a tail up to $\sim 2~\tev$, the EFT is unitary for $\zeta_{1,2} \lesssim (1 - 10)~\tev^{-4}$~\cite{Fichet:2014uka}.

Unless the new-physics scenario considered is very strongly coupled, the condition requiring to be above the monopole-pair production threshold, $M \gg \mgg$, is stronger constraint than the small $\zeta_{1,2}$. EFT typically breaks down before unitarity is violated. 

%The decoupling condition is satisfied for the allowed monopole mass values acquired in EFT \eqref{EFT-limits} and BI \eqref{bi-limit} for any magnetic charge.

As described in Section~\ref{sc:cep}, the anomalous couplings $\zeta_{1,2}$ have been constrained in $pp$ collisions of 103~\ifb of integrated luminosity by the CMS-TOTEM Collaboration at a centre-of-mass energy of 13~\tev~\cite{TOTEM:2023ewz} in a search for exclusive photon production on this effective extension of the SM. The observed elliptical allowed region at 95\% confidence level (CL) on the $(\zeta_1, \zeta_2)$ plane are expressed in an analytical form as
\begin{align}\label{ellipse}
a_0 &\zeta_1^2 + a_1 \zeta_1 \zeta_2 + a_2 \zeta_2^2 < 1, \\
&a_0 = 194.0~\tev^8, \nonumber \\
&a_1 = 161.7~\tev^8, \nonumber \\
&a_2 = 44.47~\tev^8. \nonumber
\end{align}
These limits ignore the interference between the SM and BSM amplitudes; this is justified since in this kinematic region ($\mgg>350~\gev$) the SM light-by-light cross section is negligible~\cite{Fichet:2014uka,TOTEM:2023ewz}.
From this constraint on EFT coefficients, one can derive charge- and spin-dependent mass limits for electrically and magnetically charged objects (cf.\ next section). 

%%%%%%%%%%%%%%%%%%%%%%%%%%%%%%%%%%%%%%%%%%%%%%%%
\subsection{High-electric-charge objects}\label{sc:heco}

Quartic photon couplings can be modified via loops of new heavy particles or new resonances that couple to photons~\cite{Fichet:2013gsa,Fichet:2014uka}. For energies much smaller than the mass of these exotic particles, the low-energy EFT results in higher order of the Heisenberg and Euler theory~\cite{Heisenberg:1936nmg} of LbL scattering, where $\zeta_{1,2}$ are expressed as~\cite{Fichet:2014uka}
 \begin{equation}\label{zeta}
 \zeta_{i} = \alpha^2 \qeff^4 M^{-4} c_i, \quad i=1, 2,
\end{equation}
where $\qeff^4=\sum_i Q_i^4$ for particles $i$ running in the loop, of electric charge $Q_i$ with approximately the same mass $M$. This way all additional multiplicities such as colours or flavour are accounted for, e.g.\ three for coloured particles. For the LHC centre-of-mass energies considered here, we take the value of $\alpha$ at the $Z$-boson mass scale, i.e.\ $\alpha = 1/128$, as it corresponds to the hard-scattering energy. The coefficients $c_i$ are spin-dependent:
\begin{equation}
c_1 =
\begin{cases}
  \frac{1}{288},  &s=0 \\
      - \frac{1}{36}, &s=\frac{1}{2}\\
   - \frac{5}{32},  &s=1
\end{cases}, \hspace{4mm}
c_2 =
\begin{cases}
  \frac{1}{360},  &s=0 \\
       \frac{7}{90},  &s=\frac{1}{2}\\
    \frac{27}{40,}  &s=1
\end{cases}.\label{EFT-c}
\end{equation}
Similar expressions have been derived for neutral and for spin-2 particles~\cite{Fichet:2013gsa}.

Here, we constrain HECO models indirectly via their loop contributions, as expressed by the $\zeta_{1,2}$ coefficients of \eqref{zeta}, with $c_{1,2}$ given in \eqref{EFT-c}. The  experimental constraints \eqref{ellipse} obtained from CMS-TOTEM~\cite{TOTEM:2023ewz} translate into the bounds 
\be
\label{ellipse-HECO}
\left( \frac{M}{\qeff}\right)^8 > \alpha^4 \left(  a_0 c_1^2 + a_1 c_1 c_2 +a_2 c_2^2 \right)
\:\Longrightarrow\:
\frac{M}{\qeff} > 
\begin{cases}
  44.7~\gev,  &s=0 \\
  63.2~\gev, &s=\frac{1}{2}\\
  114.5~\gev.  &s=1
\end{cases}.
\ee

For this result to be within the EFT validity requirements, $M \gg m_{\gamma\gamma,\text{max}} \Rightarrow M \gtrsim 10~\tev$. Under this assumption, the constraints \eqref{ellipse-HECO} are meaningful only if the effective electric charge satisfies: $\qeff \gtrsim 90$ for $s=0$, $\qeff \gtrsim 160$ for $s=\hf$, and $\qeff \gtrsim 220$ for $s=1$. For charges below these values, the mass range excluded by~\eqref{ellipse-HECO} lies entirely below the EFT validity threshold and ad-hoc form factors are required to reach reliable results~\cite{Fichet:2014uka}. This observation highlights the complementarity of direct and indirect searches: direct production searches, which do not rely on the EFT framework, retain their full sensitivity in the low mass regime where the (indirect) EFT approach breaks down. The excluded HECO parameter space in the $(\qeff,M)$ plane is illustrated in Fig.~\ref{fig:heco} for all three spin hypotheses. The exclusion boundary grows linearly with \qeff, since the excluded mass boundary scales as $M \propto \qeff$, as seen from \eqref{ellipse-HECO}. Consequently, for highly charged objects the indirect search provides exclusions extending to multi-\tev masses well beyond the kinematic reach of direct searches at the LHC.

\begin{figure}[htb]
    \centering
    \includegraphics[width=0.75\textwidth]{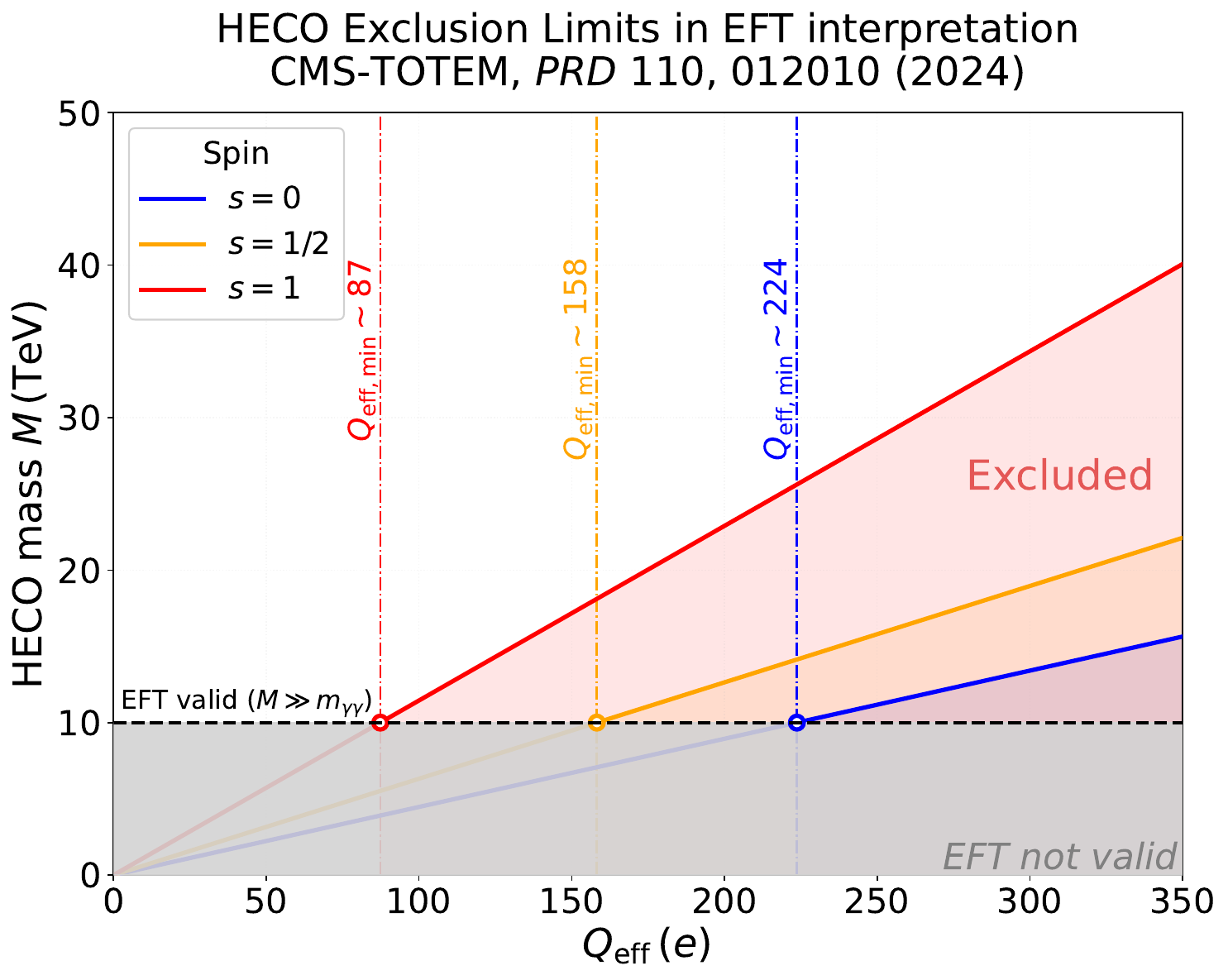}
\caption{Excluded HECO mass $M$ vs.\ effective charge \qeff for spin $s = 0$ (blue),  $s = \hf$ (orange), and $s = 1$ (red), derived from the CMS-TOTEM bounds on $\zeta_{1,2}$~\cite{TOTEM:2023ewz} via \eqref{ellipse-HECO}. Colourful shaded regions are excluded within the EFT validity domain ($M \gg \mgg$, dashed line), while the gray region below indicates masses for which EFT is not valid for the kinematic region considered experimentally. Open circles mark the minimum charge $Q_\text{eff,min}$ for which the EFT approach can be applied. In the whole EFT-allowed region, the HECO--photon (and $Z$, where applicable) coupling is too large to be treated perturbatively. }
    \label{fig:heco}
\end{figure}

A subtle point in this approach, present in all processes involving high electric or magnetic charges, is that the HECO--photon coupling is too large to be treated perturbatively. However, recently the application of tree-level Dyson--Schwinger resummation techniques led to the calculation of reliable cross sections for spin-\hf~\cite{Alexandre:2023qjo} and spin-0~\cite{Alexandre:2024pbs} HECOs. By incorporating these resummation effects, the expression for $\zeta_i$ given in \eqref{zeta} is modified as follows:
\begin{equation}\label{zeta_resum}
\zeta^{\star}_{i} = {\alpha^\star}^2 (Z^\star \qeff)^4 \mathcal{M}(\Lambda,\qeff)^{-4} c_i, \quad i=1, 2,
\end{equation}
where $Z^{\star}$ denotes the wave-function renormalisation at the ultraviolet (UV) fixed point, and $\alpha^\star = \frac{\alpha}{1+\omega^\star}$ represents the effective running coupling evaluated at the same fixed point. The quantity $\mathcal{M}(\Lambda,\qeff)$ corresponds to the running HECO mass, while $\Lambda$ is interpreted as a natural UV cutoff within this framework. Thus, $\mathcal{M}(\Lambda,\qeff)$ is identified as the physical mass $M$ of the HECO. The resummation has numerical effects for charges above $\sim11e$.
In this context, the parameters $Z^\star$ and $\omega^\star$ take specific values depending on the particle spin: for spin-\hf states, $Z^\star = 1.477$ and $\omega^\star = 0.431$~\cite{Alexandre:2023qjo}, while for spin-0 states, $Z^\star  \simeq 1.29 $ and $\omega^\star \lesssim 0.11$~\cite{Alexandre:2024pbs}.

The experimental constraints \eqref{ellipse} obtained from CMS-TOTEM~\cite{TOTEM:2023ewz}, in the case of resummation, are
\be\label{ellipse-HECO_resum}
\left( \frac{M}{Z^\star \qeff}\right)^8  > (\alpha^\star)^ 4 \left(  a_0 c_1^2 + a_1 c_1 c_2 +a_2 c_2^2 \right) 
\:\Longrightarrow\:    
\frac{M}{\qeff} > 
    \begin{cases}
        54.7~\gev, & s = 0 \\
        78.0~\gev, & s = \tfrac{1}{2}
    \end{cases}.
\ee
We observe that after resummation is applied the (reliable) mass limits are more stringent than in the tree-level case, as shown in Fig.~\ref{fig:heco_exclusion_resum}. The same effect, i.e.\ enhancement of the production rate leading to stronger bounds,  was observed when resummation was applied to the Drell--Yan and photon-fusion processes~\cite{Alexandre:2023qjo,Alexandre:2024pbs}.

\begin{figure}[htb]
    \centering
    \includegraphics[width=0.75\textwidth]{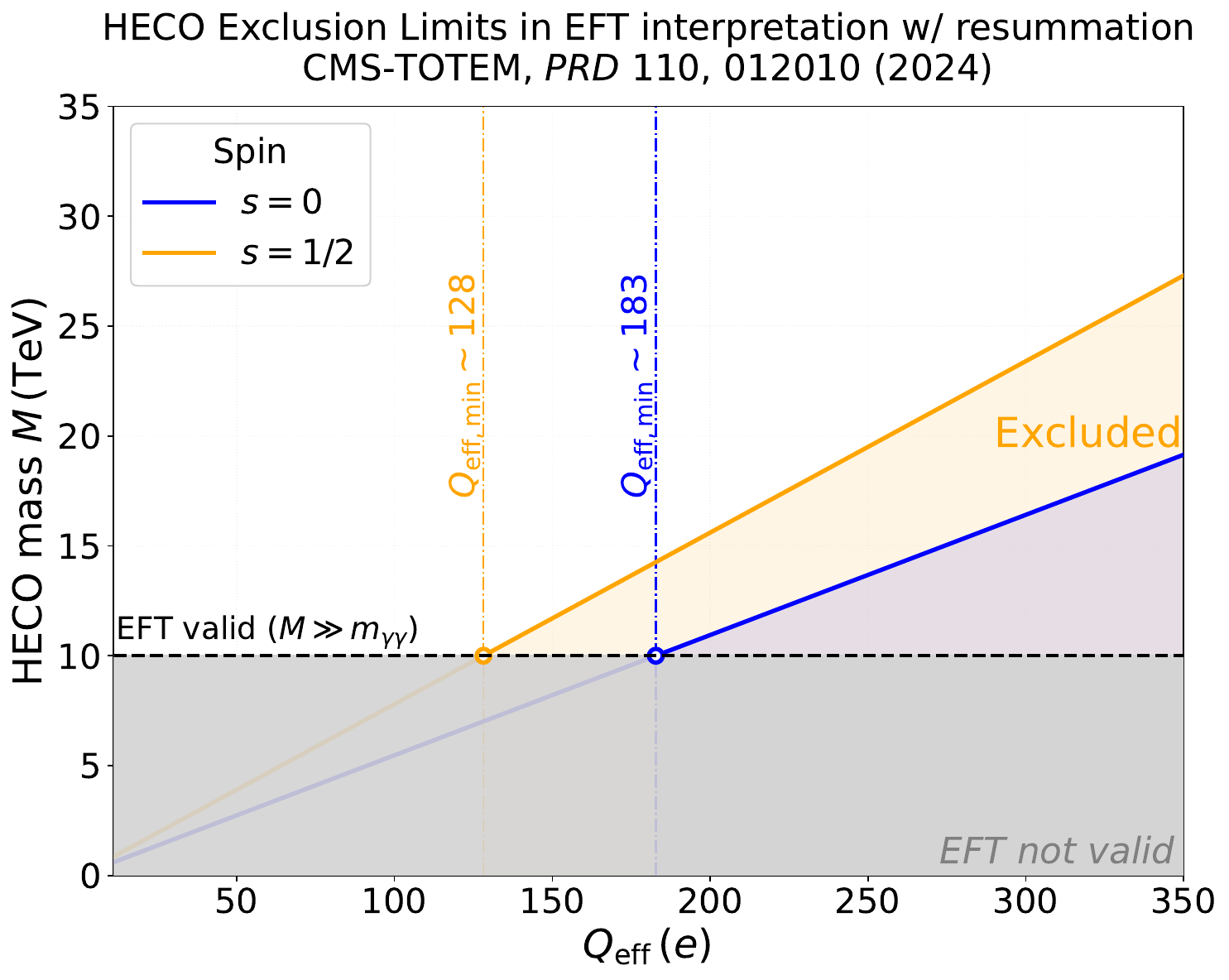}
    \caption{Excluded HECO mass $M$ vs.\ effective charge \qeff including resummation effects~\cite{Alexandre:2023qjo,Alexandre:2024pbs} for spin $s = 0$ (blue) and  $s = \hf$ (orange) derived from the CMS-TOTEM bounds on $\zeta_{1,2}$~\cite{TOTEM:2023ewz} via \eqref{ellipse-HECO_resum}. Colourful shaded regions are excluded within the EFT validity domain ($M \gg \mgg$, dashed line), while the gray region below indicates masses for which EFT is not valid for the kinematic region considered experimentally. Open circles mark the minimum charge $Q_\text{eff,min}$ for which the EFT approach can be applied.}
    \label{fig:heco_exclusion_resum}
\end{figure}

%%%%%%%%%%%%%%%%%%%%%%%%%%%%%%%%%%%%%%%%%%%%%%%%
\subsection{Magnetic monopoles}\label{sc:mm-eft}

Now, in the case of MMs, the EFT Lagrangian \eqref{EFT} is written in the form~\cite{Ginzburg:1999ej} 
\begin{equation}\label{EFT-MM}
\mathcal{L}_{4\gamma}^\text{MM} = 
\frac{1}{36} \left( \frac{g}{\sqrt{4\pi}M} \right)^4  \left[ \frac{\beta_+ + \beta_-}{2} (F_{\mu\nu}F^{\mu\nu})^2 + \frac{\beta_+ - \beta_-}{2} (F_{\mu\nu}\tilde{F}^{\mu\nu})^2 \right],  
\end{equation}
where $\tilde{F}^{\mu\nu}=\epsilon^{\mu\nu\alpha\beta}F_{\alpha\beta}/2$ is the dual field tensor of $F^{\mu\nu}$, $\epsilon^{\mu\nu\alpha\beta}$ is the four-dimensional Levi--Civita symbol, $g=n\gd$ is the MM magnetic charge from \eqref{dqc}, and the $\beta_\pm$ coefficients are:
\begin{equation}
\beta_+ =
\begin{cases}
  \frac{1}{5},  &s=0 \\
 \frac{11}{10}, &s=\frac{1}{2}\\
  \frac{63}{5},  &s=1
\end{cases}, \hspace{4mm}
\beta_- =
\begin{cases}
  \frac{3}{20},  &s=0 \\
    -\frac{3}{10},  &s=\frac{1}{2}\\
    \frac{9}{20,}  &s=1
\end{cases}. \label{EFT-beta}
\end{equation}

We define the factor $A \equiv \frac{1}{72} \left( \frac{\gd}{\sqrt{4\pi}} \right)^4 = \frac{1}{1152\alpha^2}$, using \eqref{dqc}. The value $\alpha = 1/128$ at the $Z$ mass we consider here gives $\gd = 64e$ and $A \simeq 14.21$. We note that the Dirac charge value departs from the $68.5e$ widely considered in the literature, derived from an $\alpha$ value of $1/137$ (cf.~\cite{Alexandre:2026auj}). The latter value is suitable for the photon flux calculation in EPA frameworks and quasi-real photons~\cite{dEnterria:2025ecx}.

Through the redefinitions $b_\pm$ of the coefficients $\beta_\pm$:
\begin{equation}\label{b-coeff}
b_\pm \equiv A \left( \frac{n}{M} \right)^4 ( \beta_+ \pm \beta_-), 
\end{equation}
the Lagrangian \eqref{EFT} can be written in the basis \eqref{EFT-MM} with the substitutions~\cite{Ellis:2022uxv}
\begin{equation}\label{zeta-b}
\zeta_1=b_+-2b_-, \quad \zeta_2=4b_-, 
\end{equation}
which allows the monopole parameter $M/n$ to be directly constrained by the coefficients $\zeta_{1,2}$.

If $\zeta_{1,2}$ are expressed in terms of \eqref{zeta-b} and \eqref{b-coeff}, then the experimental constraint \eqref{ellipse} is transformed into lower bounds to monopole masses:
\begin{align}\label{ellipse-MM}
\left( \frac{M}{n}\right)^8 &> A^2 (a_{++} \beta_+^2 + a_{+-} \beta_+\beta_- + a_{--} \beta_-^2 ),  \\
a_{++} &\equiv a_0 - 4a_1 + 16a_2 = 259.0~\tev^8, \nonumber \\
a_{+-} &\equiv -6a_0 + 16a_1 -32a_2 = -0.32~\tev^8, \nonumber \\
a_{--} &\equiv 9a_0 -12a_1 + 16a_2 = 517.5~\tev^8. \nonumber
\end{align}

Finally, by using the spin-specific $\beta_\pm$ values from \eqref{EFT-beta}, the constraints \eqref{ellipse-MM} lead to the following indirect limits for Dirac (i.e.\ point-like) monopoles of mass $M$,  charge $n\gd$ and spin $s$:
\begin{equation}\label{EFT-limits}
\frac{M}{|n|}  >
\begin{cases}
  2.86~\tev,  &s=0 \\
  4.05~\tev, &s=\frac{1}{2}\\
  7.33~\tev.  &s=1
\end{cases}.
\end{equation} 

These indirect exclusion limits at 95\% CL on MM mass as a function of the magnetic charge are also presented in Fig.~\ref{fig:eft} for scalar, fermionic and vector monopoles. The excluded-mass boundary scales linearly with the magnetic charge, so that higher-charge monopoles are excluded up to progressively larger masses. The most stringent limits are obtained for spin-1 (vector) monopoles, followed by spin-\hf (fermionic) and spin-0 (scalar), reflecting the spin-dependent coupling structure of the EFT operators. These limits only slightly depend on the $\alpha$ value; if $\alpha = 1/137$ is assumed, they increase by $\sim3.5\%$.

The dashed horizontal line at $M = 10~\tev$ indicates the EFT validity threshold $M \gg \mgg$; limits below this boundary, shown in the grey region, are outside the regime of validity of the EFT approach and are therefore not considered as reliable exclusions. The application of form factors is needed in this region. For spin-0 monopoles, meaningful limits are obtained starting from $4\gd$, with excluded masses up to $M \simeq 30~\tev$ at $10\gd$; for spin-\hf from $3\gd$, up to $M \simeq 40~\tev$ ($10\gd$); and for spin-1 from $2\gd$, up to $M \simeq 73~\tev$ ($10\gd$). MMs of charge $g=1\gd$ fall entirely within the EFT-invalid region for any spin value.

\begin{figure}[htb]
    \centering
    \includegraphics[width=0.75\textwidth]{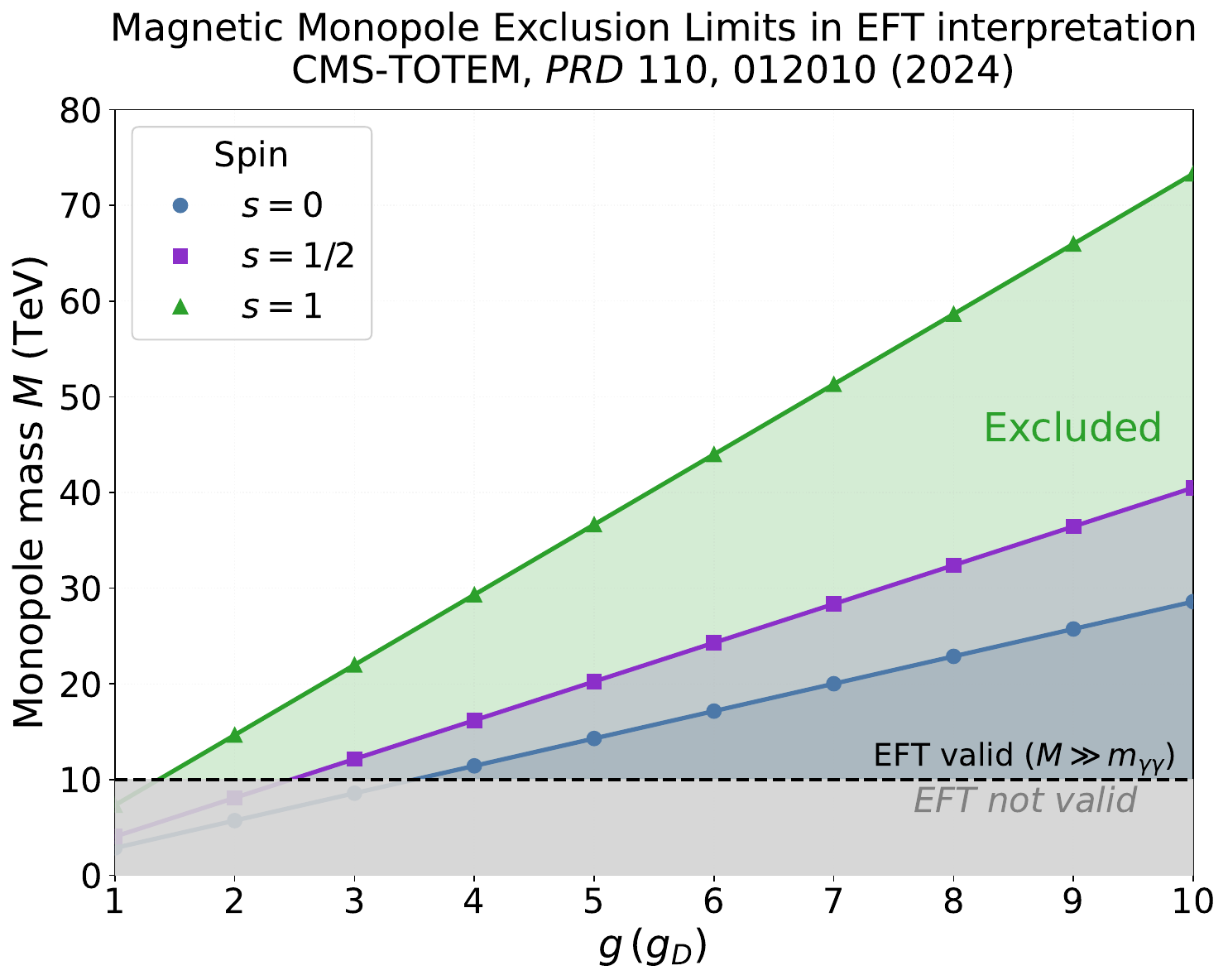}
    \caption{Exclusion lower mass limits at 95\% CL for spin-0 (blue), spin-\hf (purple) and spin-1 (green) MMs as a function of magnetic charge, $g$, in units of Dirac charge \gd, by using limits on EFT coefficients $\zeta_{1,2}$ set by CMS-TOTEM~\cite{TOTEM:2023ewz} via \eqref{EFT-limits}. The shaded regions represent the excluded parameter space within the EFT validity domain $M\gg \mgg$, indicated by the dashed line. In the gray region, EFT is not valid for the kinematic region considered experimentally. }
    \label{fig:eft}
\end{figure}

The difference in mass constraints across varying spin states stems from the coefficients $\beta_{\pm}$ listed in \eqref{EFT-beta}.  Notably, these coefficients exhibit maximal magnitudes for spin-1 monopoles and minimal values for spin-0 configurations.

As in the case of HECOs, the magnetic coupling $g$ is inherently too large to be treated perturbatively. However, very recently the application of tree-level Dyson--Schwinger resummation techniques based on~\cite{Alexandre:2019iub} justified the perturbatively obtained behaviour of monopoles~\cite{Alexandre:2026auj}.

%%%%%%%%%%%%%%%%%%%%%%%%%%%%%%%%%%%%%%%%%%%%%%%%%%%%%%%%%%%%
\section{Born--Infeld scenario}\label{sc:born}

Another approach to constrain MMs~\cite{Ellis:2017edi} is the Born--Infeld (BI) theory~\cite{Born:1934gh}. It was proposed to impose an upper bound on the electric field, by carrying out a nonlinear modification of the QED Lagrangian $\mathcal{L}_{\text{QED}}=-\frac{1}{4}F_{\mu \nu}F
    ^{\mu \nu}$ expressed by
\begin{equation}
   \mathcal{L}_{\text{BI}}= \bbi^2 \biggl(1-\sqrt{1+\frac{1}{2 \bbi^2}F_{\mu \nu}F^{\mu \nu}-\frac{1}{16\bbi^4}(F_{\mu \nu}\Tilde{F}
    ^{\mu \nu})^2} \biggr), \label{BI}
\end{equation}
where \bbi is an unknown parameter, which in four-spacetime dimensions has dimensions of mass-squared and it can be written, in general, as $\bbi \equiv M'^2$, where $M'$ is a mass scale. 

Born--Infeld regularisation has been used to create finite-energy, stable solutions~\cite{Arunasalam:2017eyu,Ellis:2016glu,Mavromatos:2018kcd,Farakos:2025byy,Mavromatos:2026nlp} for the ---initially characterised  by core singularities--- electroweak monopole. It was proposed by Cho and Maison~\cite{Cho:1996qd} as a topological soliton in electroweak theory, representing a hybrid of a Dirac monopole and a 't Hooft–Polyakov monopole. It possesses a magnetic charge of 2\gd and a mass at the \tev scale~\cite{Ellis:2016glu}, making it accessible in high-energy colliders.  
Finite-energy monopole solutions of Cho-Maison type, derived with the introduction of BI extensions of the SM hypercharge $U(1)_Y$  sector, are characterised by a mass~\cite{Ellis:2022uxv}
 \begin{equation}
M_\text{CM} = E_0 + E_1,
 \end{equation}
where $E_0$ is the contribution associated with the BI $U(1)_Y$ hypercharge, $M_Y=\cos\theta_W M'$~\cite{Arunasalam:2017eyu} with $\theta_W$ the weak mixing angle, and $E_1$ coming from the rest of the Lagrangian. Recent estimations showed that $E_0 \simeq 72.8 {M_Y}$~\cite{Arunasalam:2017eyu} and $E_1 \simeq 7.6~\tev$~\cite{Mavromatos:2018kcd,Farakos:2025byy,Mavromatos:2026nlp}. By expanding \eqref{BI} in $\bbi^{-2}$, a connection is made between  the anomalous gauge quartic couplings $\zeta_{1,2}$ of \eqref{EFT} and the BI parameter $\bbi$~\cite{Ellis:2022uxv}:
\begin{equation}\label{zeta12} 
   \zeta_1=-\frac{1}{32\bbi^2}, \quad \zeta_2=\frac{1}{8\bbi^2}. 
\end{equation}

Thus the constraints \eqref{ellipse} set by CMS-TOTEM~\cite{TOTEM:2023ewz} can be translated into a limit on $\bbi$:
\begin{equation}\label{beta-limit} 
   \bbi = M'^2 > 0.71~\tev^2,
\end{equation}
which in turn provides an upper mass limit for a Cho-Maison monopole:
\begin{equation}\label{bi-limit} 
   M_\text{CM} \simeq 7.6+ 72.8  \cos{\theta_W}M' > 61~\tev,
\end{equation}
It is reminded that, unlike the mass bounds \eqref{beta-limit}, valid for any monopole of the specified spin and charge, the limit \eqref{bi-limit} only applies to the electroweak monopole of Cho-Maison~\cite{Cho:1996qd}. This value is clearly beyond the reach of any current collider. 

%%%%%%%%%%%%%%%%%%%%%%%%%%%%%%%%%
\section{Conclusions and outlook}\label{sc:conclu}

Investigating the existence of magnetic monopoles and high-electric-charge objects is among the main open questions in particle physics. Here we constrained indirectly the MM and HECO mass based on measurements on diphoton events at the LHC through their hypothetical presence in the $\gamma \gamma \to \gamma \gamma$ box-diagram. One of the most advantageous processes having diphoton final state signature is the central exclusive $\gamma \gamma$ production in which photons are measured in a central detector and outgoing intact hadrons are tagged with dedicated forward detectors. This leads to obtain a large background rejection factor.

Indirect constraints on magnetic monopoles and HECOs are set through EFT and BI ---the latter for MMs--- interpretations, which are subject to the pertinent approximations and validity conditions. Considering EFT theories, lower bounds on MM and HECO masses have been obtained by interpreting constraints on anomalous quartic gauge couplings set by the CMS-TOTEM collaboration from a search for central exclusive diphoton production~\cite{TOTEM:2023ewz}.

For HECOs, the EFT validity requirement similarly restricts the self-consistent exclusion to high electric charges, $\qeff \gtrsim 90$, 160 and~220 for spins~0, \hf and~1 respectively, beyond which the excluded mass scales linearly with~\qeff up to multi-\tev values. Resummation schemes have been applied to treat the large HECO-photon coupling, leading to stronger bounds for $\qeff \gtrsim 130~\gev$ ($\qeff \gtrsim 180~\gev$) for scalar (fermionic) HECOs.

For MMs with spins~0, \hf and~1, exclusion limits within the EFT validity regime, $M \gtrsim 10~\tev$ are obtained for magnetic charges $\geq 4\gd$, 3\gd and 2\gd, respectively, with mass bounds reaching up to $\sim$30, 40 and 73~\tev at $g = 10\gd$. 
In the Born--Infeld context, a lower limit on the monopole mass has been set to 61~\tev. Interpretations of these precision measurements act complementary to direct searches for monopoles currently carried out by the MoEDAL and ATLAS experiments at the LHC. 

Furthermore, light-by-light scattering in heavy-ion collisions, observed at the LHC by ATLAS and CMS, opens another window to probe BSM contributions to it, including those by MMs and HECOs. In Pb--Pb collisions, the $\gamma$ spectrum is softer that in $pp$ collisions, e.g., for a centre-of-mass energy of 5.5~\tev, each photon can reach a maximum energy of $\sim80~\gev$~\cite{Shao:2022cly}. Nonetheless, the photon flux is immensely enhanced by a factor $Z^4$, with $Z$ the atomic number of the ion, with respect to the $pp$ beams. Other advantages of the heavy-ion beams are the low QCD background and lower pile-up. The measurements of this process performed by the ATLAS~\cite{ATLAS:2017fur,ATLAS:2019azn,ATLAS:2020hii} and the CMS~\cite{CMS:2018erd,CMS:2024bnt} experiments in ultraperipheral Pb--Pb  collisions can be deployed to constrain further the presence of HECOs and monopoles in loop diagrams~\cite{Baltz:2007kq}. 

Beyond the LHC, promising prospects exist for such precision measurements and consequent MM and HECO constraints, in future $e^+e^-$ colliders~\cite{Ellis:2022uxv}, $\gamma\gamma$~\cite{Jikia:1993tc,Ginzburg:1999ej,Ginzburg:2020koq}, and $\mu^+\mu^-$ colliders~\cite{Yang:2022ilt}. 
 
%%%%%%%%%%%%%%%%%%%%%%%%%%%%%%%%%%%%%%%%%%%%%%%%
\subsection*{Acknowledgements}

EM acknowledges support from the U.S. National Science Foundation under grant No.\ 2309505 (FAIN), awarded to the University of Alabama MoEDAL group for the project ``Searching for Magnetic Monopoles and Other Exotics with MoEDAL''. The research of VAM and EM is supported by the Generalitat Valenciana via the Excellence Grant Prometeo CIPROM/2021/073, by MICIN/AEI/10.13039/501100011033/FEDER, EU via the grants PID2021-122134NB-C21 and PID2024-158190NB-C21 and by MICIU/AEI grant Severo Ochoa CEX2023-001292-S. VAM acknowledges support from CSIC through grant 2025AEP129.

%%%%%%%%%%%%%%%%%%%%%%%%%%%%%%%%%%%%%%%%%%%%%%%%
\subsection*{Data Availability}
No new or original datasets were created in this study.

%%%%%%%%%%%%%%%%%%%%%%%%%%%%%%%%%%%%%%%%%%%%%%%%

%\bibliographystyle{jhep.bst}
\bibliographystyle{JHEP-vaso}
\bibliography{sample}

\end{document}